\documentclass[twocolumn,pra,preprintnumbers,amsmath,amssymb,superscriptaddress,floatfix]{revtex4}

\usepackage{latexsym}
\usepackage{epsfig}
\usepackage{graphicx}
\usepackage{textcomp}
\usepackage{color}

\setcitestyle{super}
\begin{document}

\title{Multi-state discrimination below the quantum noise limit at the single-photon level}

\author{A. R. Ferdinand}

\affiliation{Center for Quantum Information and Control,
1 University of New Mexico, Albuquerque, New Mexico 87131-0001}

\author{M. T. DiMario}

\affiliation{Center for Quantum Information and Control,
1 University of New Mexico, Albuquerque, New Mexico 87131-0001}

\author{F. E. Becerra}

\affiliation{Center for Quantum Information and Control,
1 University of New Mexico, Albuquerque, New Mexico 87131-0001}
\email{fbecerra@umd.edu}

%

\begin{abstract}

Measurements approaching the ultimate quantum limits of sensitivity
are central in quantum information processing, quantum metrology, and
communication. Quantum measurements to discriminate multiple states
at the single-photon level are essential for optimizing information
transfer in low-power optical communications and quantum
communications, and can enhance the capabilities of many quantum
information protocols. Here, we theoretically investigate and
experimentally demonstrate the discrimination of multiple coherent
states of light with sensitivities surpassing the quantum noise
limit (QNL) at the single-photon level under realistic conditions of
loss and noise based on strategies implementing globally-optimized
adaptive measurements with single photon counting and displacement
operations. These discrimination strategies can provide realistic advantages to enhance information transfer at
low powers, and are compatible with
photon number resolving detection, which provides robustness at high
powers, thus allowing for surpassing the QNL at arbitrary input
power levels under realistic conditions.
\end{abstract}

\pacs{03.67.Hk, 03.65.Ta,  42.50.Ex} \maketitle

\noindent
\textbf{INTRODUCTION}
\\
Assessing the information about the state of a quantum system is
fundamentally limited by quantum noise. This task requires
performing measurements discriminating among different states, and
when these states are nonorthogonal, discrimination cannot be
performed perfectly due to their intrinsic overlap
\cite{helstrom76}. Measurements for the discrimination of
nonorthogonal states, such as coherent states, can enable
secure communications \cite{huttner95, bennett92,
grosshans02, gisin02, weedbrook12,sych10,leverrier09,leverrier11, qi15, soh15}
and assist quantum information protocols for quantum repeaters
\cite{vanloock06,vanloock08}, quantum computing
\cite{munro05,nemoto04,ralph03}, entanglement generation with high
fidelity \cite{kilin11, vanloock11}, quantum signatures
\cite{collins14, croal16}, and quantum finger printing
\cite{arrazola14, guan16}. Moreover, efficient discrimination
measurements of multiple coherent states can enhance information
transfer in communication \cite{becerra15, rosati16, lee16, tan15},
and, by performing collective measurements over long sequences of
states, allow for achieving the ultimate limits for classical
information transfer (classical capacity \cite{caves94}) in
communication channels with loss and noise \cite{giovannetti04,
giovannetti14, mari14}.

Quantum mechanics allows for measurements that in principle achieve
the ultimate sensitivity limits for the discrimination of
nonorthogonal states \cite{helstrom76, yuen75}. These measurements can optimize
many quantum information and communication protocols. The lowest
error probability allowed by quantum mechanics for discriminating
nonorthogonal states, such as coherent states with different phases,
is referred to as the Helstrom bound \cite{helstrom76}. This bound
is much lower than the limit of ideal conventional
measurements, referred to as the quantum noise limit (QNL), which
can be reached by the homodyne (heterodyne) measurement
for two (multiple) phase states \cite{weedbrook12}. Strategies for the discrimination
of two coherent states below the QNL have
been investigated theoretically \cite{kennedy72,taeoka08,dolinar73}
and demonstrated experimentally \cite{cook07, wittman08, tsujino11},
including the optimal strategy saturating the Helstrom bound
\cite{dolinar73} based on photon counting and real-time feedback
\cite{cook07}. The discrimination of multiple coherent states below
the QNL was first investigated by Bondurant \cite{bondurant93}
generalizing the optimal two-state strategy in \cite{dolinar73} to
four states. Further theoretical \cite{nair12,izumi12,izumi13,li13,
muller15} and experimental work demonstrated that the discrimination
of multiple states below the QNL \cite{muller12} can be achieved
with current technologies \cite{becerra13}, and that some realizable
measurements show the same exponential scaling as the Helstrom bound
in the limit of high powers \cite{bondurant93, izumi12, nair12}.
Recent work showed that photon-number resolution (PNR) provides
robustness against realistic noise for discrimination of multiple
states \cite{izumi13,li13}. And an experimental realization with
detectors with finite PNR demonstrated multiple-state discrimination
below the QNL at high input-power levels, which provides advantages
for coded communication over heterodyne measurements
\cite{becerra15}.

Discrimination of multiple states surpassing the QNL in the
single-photon regime has a large potential for applications in
quantum information
\cite{vanloock06,vanloock08,munro05,nemoto04,ralph03} and enhancing
information transfer in low-power communication \cite{becerra15,
rosati16, lee16, tan15}. Moreover, optimized measurements for
nonorthogonal states at
the single-photon level can potentially be used to enhance the rate
of secure communications in quantum key distribution (QKD) with
coherent states \cite{wittmann10, leverrier09,leverrier11, sych10}, which
require states at low powers to ensure security.
However, even though theoretical advancements have shown that this
discrimination task is possible \cite{muller15}, the discrimination
of multiple states surpassing the QNL at the single-photon level under realistic
conditions with noise and loss has not been experimentally
demonstrated.

\begin{figure}[htbp]
\centering\includegraphics[width=8.5cm]{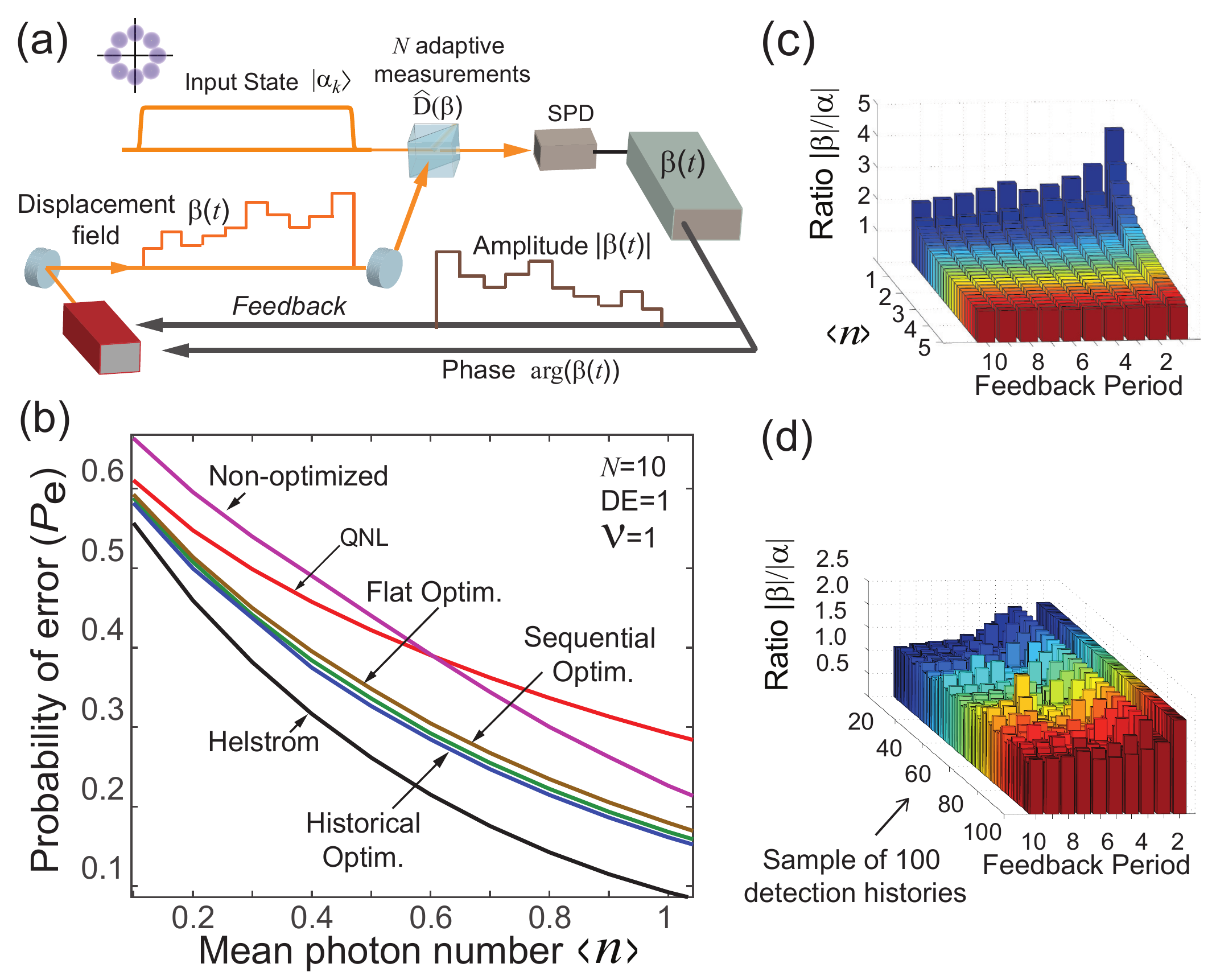}
\caption{\label{TheoryFig1} \textbf{Optimized discrimination of
multiple states.} (a) Discrimination of an input coherent state
$|\alpha_{k}\rangle\in\{|e^{i\phi_{k}}|\alpha_{k}|\rangle\}$ where
$\phi_{k}=\frac{2\pi k}{M}$ $(k=1,2,..,M)$ implementing $N$ adaptive
measurements based on optimized displacement operations
$\hat{D}(\beta)$ and single-photon counting with fast feedback. This
strategy discriminates multiple states below the quantum noise limit
(QNL) at arbitrary input power levels with a small number of
adaptive measurements. (b) Probability of error for the
discrimination of four nonorthogonal states $\{|\alpha\rangle,
|i\alpha\rangle,|-\alpha\rangle,|-i\alpha\rangle\}$ with three
different optimized strategies: flat optimization, sequential
optimization, and historical optimization, as a function of the mean
photon number $\langle n\rangle=|\alpha|^2$ based on $N=10$
optimized adaptive measurements with perfect detection efficiency
DE=1 and visibility $\mathcal{V}=100\%$ of the displacement
operation. The QNL and the Helstrom bound are shown for reference,
as well as the non-optimized strategy. (c,d) Examples of ratios
$|\beta|/|\alpha|$ of the displacement field $|\beta|$ to input
amplitude $|\alpha|$ for optimized strategies corresponding to (c)
sequential optimization for $\langle n\rangle$ from 0.2 to 5 and
(d) historical optimization for 100 possible detection histories $D_{H}$. These
ratios are the result of global optimizations that minimize the
probability of error (see main text for details). SPD, single-photon
detector; M, mirror.}
\end{figure}

Here, we theoretically investigate and experimentally demonstrate
the discrimination of multiple nonorthogonal states at the
single-photon level below the QNL based on adaptive measurements,
globally optimized displacement operations, and photon counting.
These strategies are compatible with PNR detection to extend
discrimination to high powers \cite{becerra13, becerra15}, and thus
enable discrimination strategies of multiple states below the QNL at
arbitrary input-power levels under realistic conditions. These
optimized discrimination measurements more closely approach the quantum
limits of detection for multiple states at the single-photon level,
and can be used to optimize quantum information protocols assisted
by coherent states, and enhance
information transfer in optical communication. We show that these
optimized measurements can provide advantages for increasing
information transfer with coherent states at low powers \cite{lee16} beyond what
can be achieved with the heterodyne measurement at the QNL. We
expect that these measurements can be used
in joint-detection schemes over sequences of coherent states for approaching the quantum
limits for information transfer at the single-photon level
\cite{rosati16}.
\\\\
\noindent
\textbf{RESULTS}
\\
\textbf{Optimized discrimination measurements} \\
Figure \ref{TheoryFig1}(a) shows the measurement scheme for the
discrimination of multiple nonorthogonal states
$\{|\alpha_{k}\rangle\}$ below the QNL at the single-photon level
implementing $N$ adaptive measurements based on displacement
operations $\hat{D}(\beta)$ optimized to minimize the probability of
error ($P_{\textrm{e}}$). This optimization gives rise to discrimination
strategies surpassing the QNL at arbitrary input power levels under
realistic noise and loss requiring just a few adaptive measurements,
while being compatible with PNR, which increases robustness at high
powers \cite{becerra15}. Compared to previous strategies described
in Ref. \cite{muller15} that require measurements with infinite
feedback bandwidth, the optimized strategies discussed here are
based on a finite number of adaptive measurements, which is
practical and allows for the demonstration of multi-state
discrimination beyond the QNL at the single-photon level with
current technologies.
\normalsize

The optimized strategies implement globally optimized displacements
$\hat{D}(\beta)$ of the input state $|\alpha_{k}\rangle$ from a set
of states with \footnote{This information encoding
scheme is referred to as M-ary phase-shift keying (M-PSK) in coherent communication.}
$|\alpha_{k}\rangle\in\{|e^{i\phi_{k}}|\alpha_{k}|\rangle\}$
($\phi_{k}=\frac{2\pi k}{M}$ $(k=1,2,..,M)$) to the vacuum state
$|0\rangle$, single-photon counting, and recursive Bayesian
inference. In a given adaptive measurement $j$ ($j=1,2,..,N$), the
displacement of the input state $|\alpha_k\rangle$ results in state
$|\psi\rangle=\hat{D}(-\beta)|\alpha_k\rangle$, which is followed by
photon counting with probability of $n$ photon detection given by
$\textbf{P}(n|\alpha_{k},\beta)=|\langle
n|\hat{D}(-\beta)|\alpha_{k}\rangle|^2= \frac{ \bar{n} ^{n}}{
n!}e^{-\bar{n}}$. Here $\bar{n}= |\alpha|^2 + |\beta|^2 -
2\mathcal{V}|\alpha||\beta| \cos[{\arg(\alpha_{k}) - \arg(\beta)}]$
is the mean photon number of the state $|\psi\rangle$ and
$\mathcal{V}$ is the visibility of the displacement operation
\cite{becerra13,becerra15}. After an adaptive-measurement period
$j$, the strategy estimates the posterior probabilities of the
possible input states $\{|\alpha_{k}\rangle\}$ in $j$ based on the
photon detection result $d_{j}\in\{0,1,...\}$ and the displacement
field $\beta$ as:
\begin{equation}
\footnotesize{
\emph{\textbf{P}}_{\textrm{post}}(\{|\alpha_{k}\rangle\}|\beta,d_{j})=
\frac{\textbf{P}(d_{j}|\{|\alpha_{k}\rangle\},
\beta)~\emph{\textbf{P}}^{j}_{\textrm{prior}}(\{|\alpha_{k}\rangle\})}{\sum_{\{|\alpha_{k}\rangle\}}\textbf{P}(d_{j}|\{|\alpha_{k}\rangle\},
\beta)~\emph{\textbf{P}}^{j}_{\textrm{prior}}(\{|\alpha_{k}\rangle\})},
}
\end{equation}
\\
where
$\emph{\textbf{P}}_{\textrm{post}}~(\{|\alpha_{k}\rangle\}|\beta,d_{j})$
and $\emph{\textbf{P}}^{j}_{\textrm{prior}}(\{|\alpha_{k}\rangle\})$
are the posterior and prior probabilities in $j$, respectively, for
all possible input states $\{|\alpha_{k}\rangle\}$, and
$\sum_{\{|\alpha_{k}\rangle\}}$ indicates the sum over all states
$\{|\alpha_{k}\rangle\}$. The discrimination strategy uses recursive
Bayesian inference to update prior probabilities
$\emph{\textbf{P}}_{\textrm{prior}}(\{|\alpha_{k}\rangle\})$ in
subsequent adaptive measurements, so that the posterior probability
estimated in adaptive measurement $j$,
$\emph{\textbf{P}}_{\textrm{post}}~(\{|\alpha_{k}\rangle\}|\beta,d_{j})$,
is defined as the  prior probability
$\emph{\textbf{P}}^{j+1}_{\textrm{prior}}(\{|\alpha_{k}\rangle\})$
for the next adaptive measurement, $j+1$. The phase of the
displacement field $\beta$ in $j+1$ is chosen to test the most
likely state based on the maximum-posterior probability criterion:
the phase of $\beta$ is set equal to the phase of the state for
which $P_{post}$ is maximum. This criterion for choosing the phases
of  $\beta$ is applied to all the optimized strategies described
below.

The overall probability of error $P_{e}$ after the last adaptive
measurement $N$ can be expressed as:
\begin{equation} \label{ProbError}
P_{e}=1-\sum_{D_{H}}P_{D_{H}}\max_{\{\alpha_{k}\}}\{P^{N}_{\textrm{prior}}(\{\alpha_{k}\})P(d_{N}|\{\alpha_{k}\},[\beta])\}
\end{equation}
\\
where $[\beta]$ is the set of all the displacement fields used in
all the adaptive measurements $j=1,..,N$, which can be optimized to
minimize the final probability of error. $\max_{\{\alpha\}}$ takes
the maximum over input states  $\{|\alpha_{k}\rangle\}$ of
$P^{N}_{\textrm{prior}}(\{\alpha_{k}\})P(d_{N}|\{\alpha\},[\beta^{opt}])$,
which is proportional to the posterior probability in $N$,  and the
sum $\sum_{D_{H}}$ is realized over all detection histories
$D_{H}=\{d_{1},d_{2},...,d_{N}\}$ with occurrence probabilities
$P_{D_{H}}$. Note that $P^{N}_{\textrm{prior}}(\{\alpha_{k}\})$ in
Eq. (\ref{ProbError}) depends on the detection histories $D_{H}$ and
the displacement fields $[\beta]$ in all the previous adaptive
measurements. The freedom to choose $[\beta]$ allows for finding global
optimizations of the set of displacement field amplitudes $[\beta]$
that minimize the overall probability of error, and gives rise to
different discrimination strategies with different complexities and
degrees of sensitivity. These optimized strategies for which
$|\beta|\neq |\alpha|$ enable measurements surpassing the QNL in the
low-power regime, a task that measurements without optimized
amplitudes (non-optimized strategies) for which $|\beta|=|\alpha|$ cannot achieve.

The simplest optimized discrimination strategy finds a constant value of the
amplitude of the displacement field $|\beta|$ over
all the $N$ adaptive measurements to minimize the probability of
error, so that
\begin{equation}\label{PeFlt}
\frac{\delta Pe}{\delta\beta}|=0.
\end{equation}
Figure \ref{TheoryFig1}(b) shows the probability of error for the discrimination
of four nonorthogonal coherent states $\{|\alpha\rangle,
|i\alpha\rangle,|-\alpha\rangle,|-i\alpha\rangle\}$ achieved
by this strategy for N=10. We observe that this strategy is
sufficient to surpass the QNL for multiple states at the
single-photon level. This strategy, which we call ``flat
optimization,'' is the discrete version of that described in
\cite{muller15} and only requires a finite $N$, which translates
to measurements with finite bandwidth and provides advantages for realistic
implementations.

Performing more sensitive measurements requires implementing global
optimization over the magnitudes of the displacement fields
$[\beta]$ in individual adaptive measurements. A simple strategy
that we refer to as ``sequential optimization'' implements $N$
optimized displacement amplitudes
$[\beta]=\{\beta_{1},\beta_{2},...,\beta_{N}\}$ one in each adaptive
measurement, but optimized simultaneously so that
\begin{equation}\label{PeSeq}
\frac{\delta Pe}{\delta\{\beta_{1,...,N}\}}=0.
\end{equation}

Figure \ref{TheoryFig1}(b) shows the improved performance of this strategy
over the ``flat optimization'' strategy. This ``sequential
optimization'' strategy is similar to the optimal discrimination
strategy for two states described by Dolinar \cite{dolinar73}, but
for multiple states. We observe that the displacement amplitude
ratio $|\beta|/|\alpha|$ starts with a large value and decreases as a
function of time, as seen in Fig. \ref{TheoryFig1}(c). However, in
contrast to the optimal measurement for two states \cite{dolinar73},
the displacement magnitude ratio is not a monotonically decreasing
function of time, and may not be optimal as
$N\rightarrow\infty$.

Higher sensitivities for discrimination of multiple states can be
achieved by strategies based on globally optimized displacement
operations that are conditional on the detection histories. These
strategies have as many optimal displacement amplitudes as the
number of possible detection histories $D_{H}$ minus one.
For detectors without PNR capabilities, i.e. with only two possible
detection outcomes, the number of detection histories is $2^N$, and
the displacement amplitudes
$[\beta]=\{\beta_{1},\beta_{2},...,\beta_{2^{N}-1}\}$ are optimized to minimize
the final probability of error
\begin{equation}\label{PeHist}
\frac{\delta Pe}{\delta\{\beta_{1,...,2^N-1}\}}=0.
\end{equation}
These strategies, which we refer to as ``historical optimization,''
generalize strategies for discrimination of multiple states based on
coherent displacements, photon counting and finite number of
adaptive measurements with maximum posterior probability criteria.

Figure \ref{TheoryFig1}(b) shows the improved performance of the historical optimization strategy
over the ``flat optimization'' and the ``sequential
optimization.'' The historical optimization is closer to the Helstrom limit for
a small number of adaptive measurements. Fig. \ref{TheoryFig1}(d) shows the highly
complex evolution of
the optimized displacement amplitudes for this strategy for 100
possible detection histories $D_{H}$ for $N=10$. It is unknown if the historical optimization strategy
would reach the Helstrom limit for $N\rightarrow\infty$. Since the number of parameters to globally optimize grows as
$2^N$, finding the optimal $[\beta]$ for large $N$ numerically becomes computationally intensive.
However,
this strategy as well as the flat and sequential optimization strategies can be realized experimentally with current
technologies. Once optimal displacements for different strategies are determined, they can be coded in a high-bandwidth electronic controller. The controller updates the values of phase and amplitudes
of the optimal displacements conditioned on detection results in each adaptive measurement using fast feedback. In this way, the
overhead in computation time for estimating the optimal $[\beta]$ is done offline, and the complexity in the implementation of different strategies becomes comparable
in the experiment. This method allows us to experimentally demonstrate discrimination of multiple
nonorthogonal states below the QNL at the single-photon level.
\\

\begin{figure}[htbp]
\centering\includegraphics[width=8.5cm]{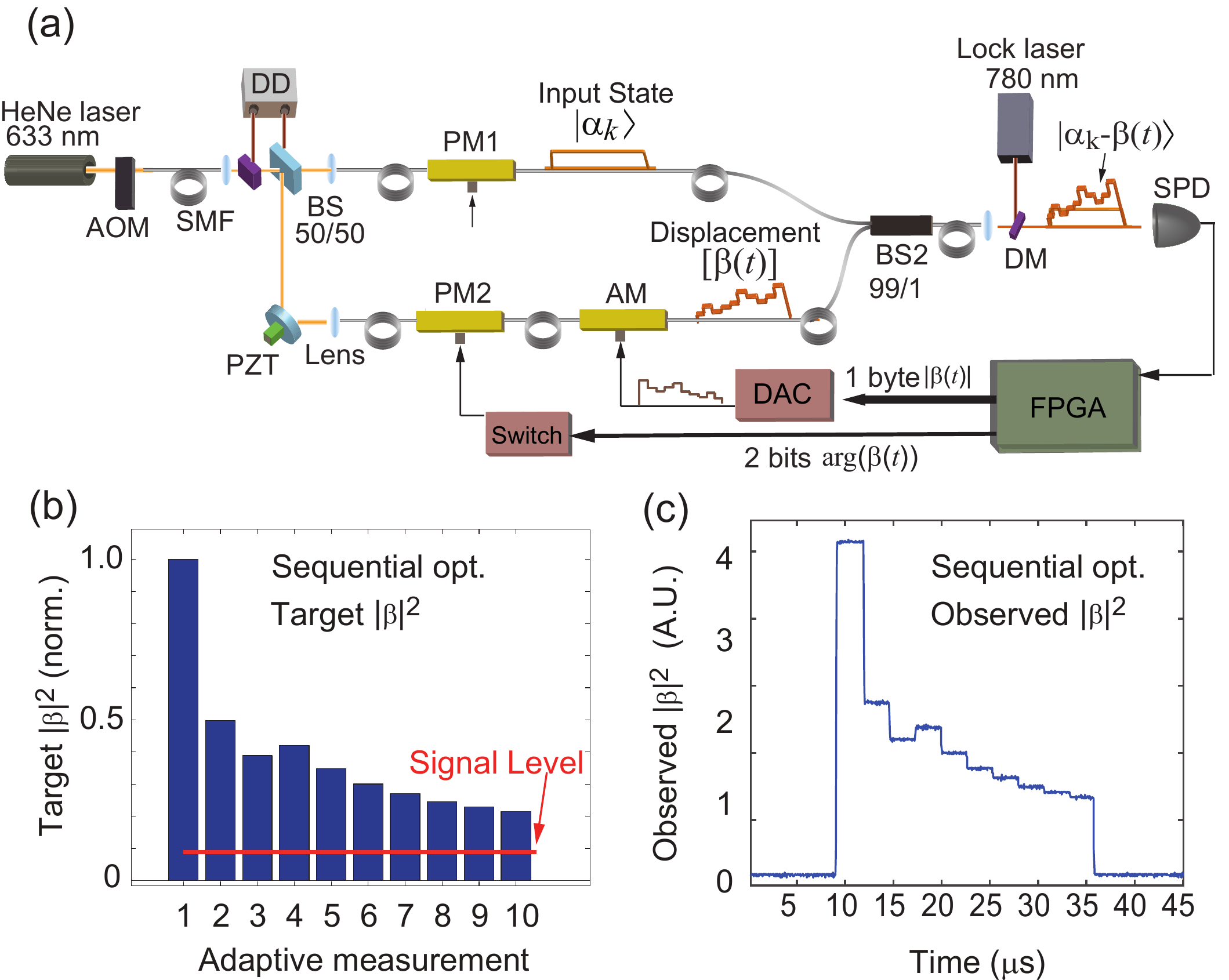}
\caption{\label{ExpConfFig} \textbf{Experimental implementation of
multi-state discrimination with optimized displacements.} (a)
Experimental configuration of strategies with optimized
discrimination of four nonorthogonal coherent states
$\{|\alpha\rangle,
|i\alpha\rangle,|-\alpha\rangle,|-i\alpha\rangle\}$ implementing
$N$=10 adaptive measurements. A phase modulator PM1 prepares the
input state, and phase PM2 and amplitude AM modulators prepare the
optimized displacements $[\beta]$. The displacement operation is
implemented by interference in a 99/1 beam splitter (BS2), and a
field-programmable gate array (FPGA) processes the detection result
from the single-photon detector (SPD) to prepare the optimized
displacements $[\beta]$ for subsequent adaptive measurements. The
phase of $[\beta]$ is controlled with a 2-bit fast multiplexer
switch and PM2, and its amplitude with a 1-byte depth
digital-to-analog converter (DAC) and the AM. SMF, single-mode
fiber; DM, dichroic mirror; PZT, piezo; AOM, acousto-optic
modulator. (b,c) Example of the intensity $|\beta|^2$ of the optimized displacement
field $[\beta]$ for the sequential global optimization strategy for
$N$=10, DE=70\%, $\mathcal{V}$=99.6\% and $\langle n\rangle=0.5$ for (b)
target and (c) observed.}
\end{figure}

\textbf{Experimental Demonstration}\\
Figure \ref{ExpConfFig}(a) shows a schematic of the experimental
demonstration of optimized strategies for the discrimination of four
coherent states $|\alpha_{k}\rangle\in\{|\alpha\rangle,
|i\alpha\rangle,|-\alpha\rangle,|-i\alpha\rangle\}$ below the QNL. A
633 nm laser with an acousto-optic modulator (AOM) prepares flat top 27 $\mu$s
long pulses defining the temporal extent of the input state $|\alpha_{k}\rangle$.
Phase modulator (PM1) prepares the phase of $|\alpha_{k}\rangle$ with a given
mean photon number $\langle n\rangle$ calibrated with a
transfer-standard calibrated detector \cite{gentile96}. Phase
(PM2) and amplitude (AM) modulators prepare the phase and the
amplitude of the optimal displacement fields $[\beta]$ for different
optimized discrimination strategies. The displacement operations are
performed by interference in a 99/1 beam splitter (BS2), achieving
an average visibility of the displacement operation of
$\mathcal{V}_{Exp}=99.6(1)\%$. An avalanche photodiode single-photon detector
(SPD) with detection efficiency of DE=$84.0(5)\%$ detects the
photons in the displaced state
$|\psi\rangle=\hat{D}(-\beta)|\alpha_k\rangle$. A field-programmable
gate array (FPGA)  processes the detection results $d_{j}$ and
prepares the optimal displacements $\hat{D}([\beta])$ for subsequent
adaptive measurements in real time with AM and PM2 for a specific
discrimination strategy (see METHODS). The experimental rate is 11 kHz with a
50$\%$ duty cycle: 50$\%$ of the time is used either for performing
the experiment or for calibration of the phase and amplitude
modulators; the other $50\%$ of the time is used for active
stabilization of the interferometer using a 780 nm laser, a
differential detector (DD) and a piezo (PZT), as in Refs.
\cite{becerra13, becerra15}. Figs. \ref{ExpConfFig}(b,c) show an example
of the target (b) and observed intensity for (c) optimized
displacement amplitudes $[\beta]$ for the sequential optimized
strategy for $\langle n\rangle=0.5$ with $N=10$. This experimental
setup achieves an overall detection efficiency of
DE$_{Exp}=70.5(7)\%$ and is compatible with PNR detection, which
increases robustness at high powers \cite{becerra15}. We use our
experiment to demonstrate globally-optimized strategies for the
discrimination of multiple coherent states below the QNL at the
single-photon level without correction for detection inefficiencies
or dark counts.

\begin{figure}[htbp]
\centering\includegraphics[width=8.5cm]{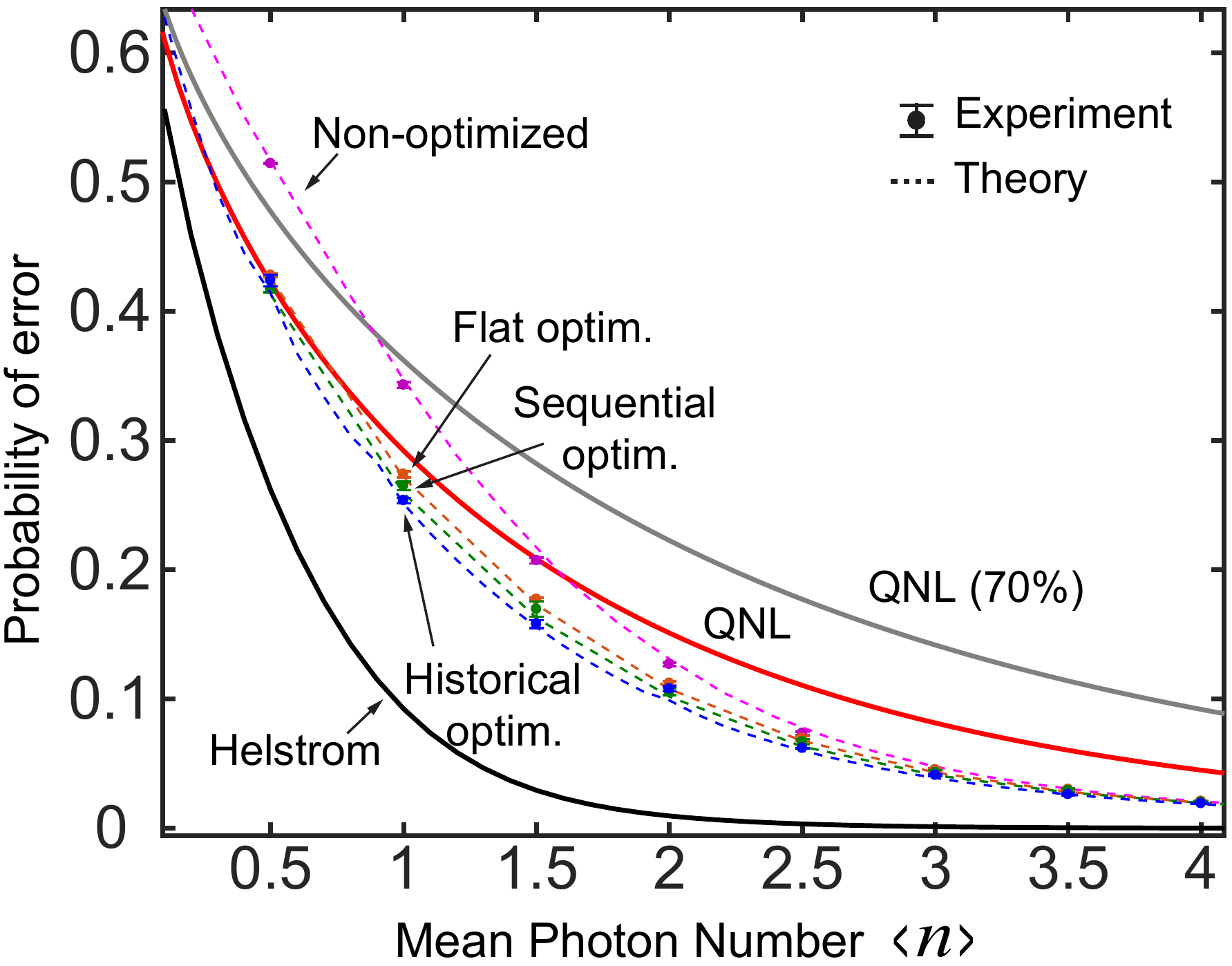}
\caption{\label{ExpResults} \textbf{Experimental results.} Error
probability for the discrimination of four nonorthogonal states
$\{|\alpha\rangle, |i\alpha\rangle, |-\alpha\rangle,
|-i\alpha\rangle\}$ with three optimized strategies, flat (light-brown dots) sequential (green
dots) and historical (blue dots) optimization, and the non-optimized
strategy (magenta dots). The Helstrom bound (black line) and the QNL
(red line) are shown for reference together with the QNL with the
same experimental conditions of detection efficiency
DE$_{Exp}$=70$\%$ (gray line). While discrimination strategies
without optimized displacements do not surpass the QNL at the
single-photon level, globally optimized strategies can reach error
rates below the QNL without any correction for detection efficiency
as shown for $\langle n\rangle=1$. Furthermore, for a system with the same
DE=70$\%$, optimized strategies surpass the QNL at arbitrary small
powers in the presence of noise and imperfections. Error bars
represent 1 statistical standard deviation from 4 runs of
1$\times10^5$ independent experiments per data point. The
theoretical predictions (dashed lines) are based on numerical
simulations with the experimentally determined detection efficiency
DE$_{Exp}$=70$\%$, visibility of the displacement
$\mathcal{V}_{Exp}=99.6\%$, and dark counts of 0.1$\%$ per pulse.}
\end{figure}

Figure \ref{ExpResults} shows the experimental results of the
discrimination of four nonorthogonal coherent states below the QNL
at the single-photon level based on three optimized strategies: flat optimization,
sequential optimization, and historical optimization. Included in the figure are
the non-optimized strategy, the Helstrom bound, the ideal QNL, and
the QNL scaled to the same system detection efficiency (70$\%$) as
in our experiment. The theoretical predictions are shown in dashed
lines and are based on numerical simulations using Eq.
(\ref{ProbError}) with the optimal displacements $[\beta]$ for
the flat, sequential, and historical
optimization strategies satisfying Eq. (\ref{PeFlt}), Eq. (\ref{PeSeq}), and Eq. (\ref{PeHist}),
respectively, with the experimental parameters DE$_{Exp}$=70$\%$,
visibility $\mathcal{V}_{Exp}=99.6\%$ and dark counts of $0.1\%$ per
pulse. We observe that the strategy based on displacements without
optimized amplitudes does not surpass the QNL at the single-photon
level, in agreement with previous work \cite{bondurant93, becerra11, becerra13,
muller12, muller15}. On the other hand, discrimination measurements
with globally optimized displacements can surpass the QNL at the
single-photon level even with moderate detection efficiency and
visibility with just 10 adaptive measurements. We observe that
sequential and historical optimizations provide similar advantages
within the experimental noise with error probabilities about 10$\%$
below the QNL at $\langle n\rangle=1$, and provide slightly greater
advantages than the flat optimization. Moreover, compared to a
system with the same DE (70$\%$), optimized measurement strategies
surpass the QNL at very small powers, which is essential for
enhancing information transfer in optical communication at low
powers beyond classical limits \cite{tan15, lee16, rosati16} and may
provide advantages for increasing security in QKD with coherent states. In addition, these
strategies can extend discrimination to high powers and increase
robustness by implementing PNR detection, which broadens their capabilities for
general tasks in quantum information and communications.

\begin{figure}[htbp]
\centering\includegraphics[width=8.5cm]{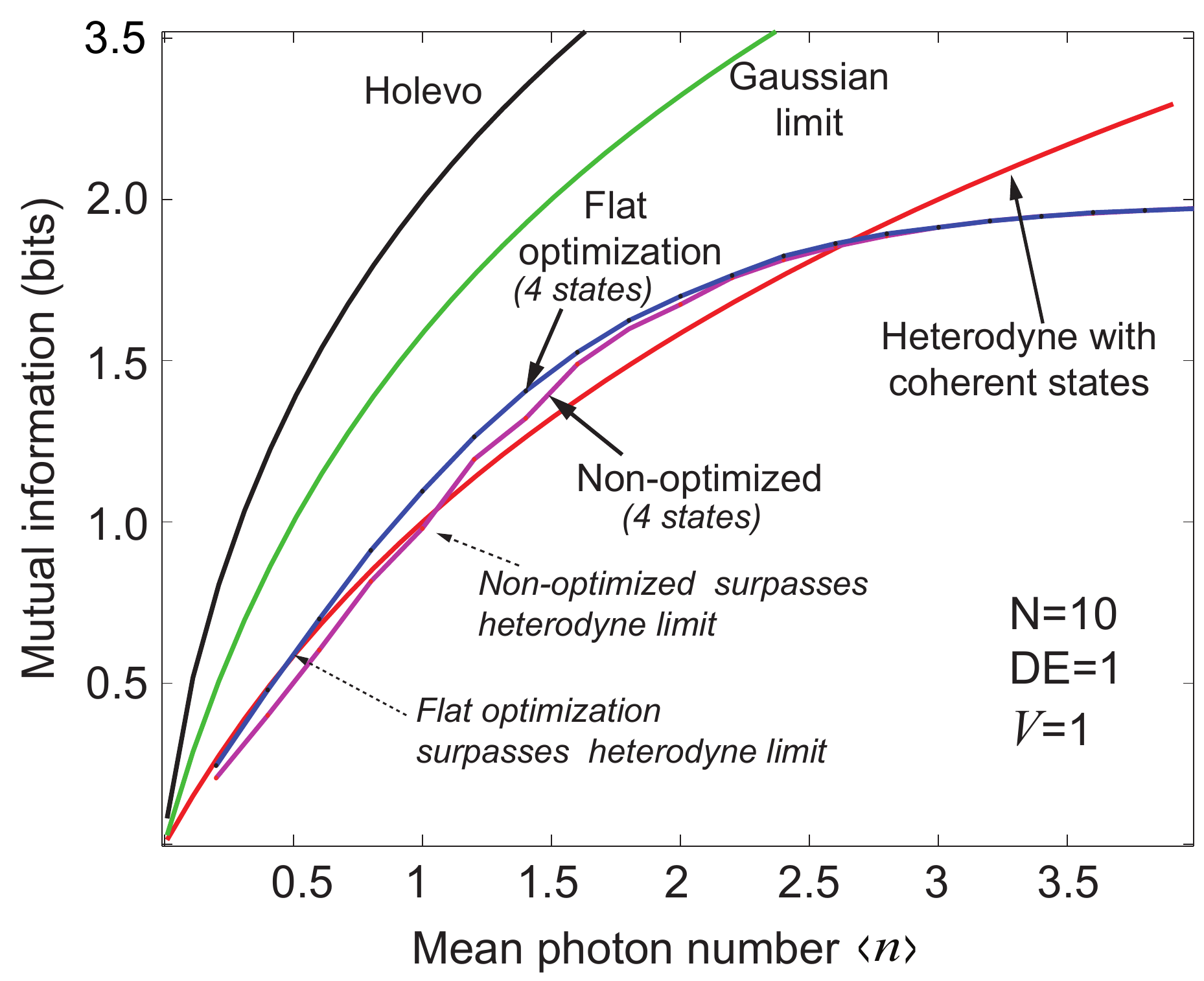}
\caption{\label{RScode} \textbf{Mutual information of optimized
discrimination measurements.} The flat optimization strategy achieves
higher mutual information than non-optimized measurements, and
surpasses the achievable capacity of heterodyne measurement over
coherent states in a range of input powers in the single-photon
regime from 0.5 to 2 photon number with just $N$=10 adaptive
measurements and four states. Other optimized measurements are
expected to provide greater advantages in capacity. The Gaussian
limit corresponds to the achievable capacity with arbitrary Gaussian states
and operations; the Holevo bound corresponds to the
ultimate limit in capacity.}
\end{figure}

\noindent
\textbf{DISCUSSION}
\\
We investigated the potential advantages of optimized strategies
for multiple states for increasing information transfer in communication.
Recent theoretical work showed that nonconventional receivers for
coherent states
based on photon counting can provide higher
communication rates than what is possible with conventional ones
\cite{taeoka14, tan15, lee16}. The optimized discrimination strategies for
multiple states presented here
can achieve higher levels of information
transfer than both non-optimized strategies and what could be achieved
with coherent states and heterodyne detection \cite{lee16} at the single photon level. Fig. 4
shows the mutual information achieved by the ``flat optimization"
discrimination measurement for four states $\{|\alpha\rangle,
|i\alpha\rangle, |-\alpha\rangle, |-i\alpha\rangle\}$ with DE=1 and
$N=10$. This optimized measurement surpasses the capacity of the heterodyne measurement using an arbitrary number
of coherent states in the single-photon regime at $\langle n\rangle=1$ \cite{lee16}, which is not possible with
non-optimized strategies. Other optimized measurements are expected to provide greater
advantages over the heterodyne measurement in the single-photon regime
However, we note that flat optimization may be easier to implement than
sequential and historical optimizations, and may be more practical in communication, albeit
with a slightly lower attainable sensitivity. This example shows the potential of
optimized measurements for low-power communication with multiple
states to increase capacity based on single-state decoding
\cite{shannon48}, which can be further optimized with larger
alphabets and optimal distribution of the input states \cite{lee16}.
Moreover, optimized discrimination  strategies can  assist
joint quantum measurements over sequences of coherent states \cite{guha11,
jarzyna16} with multiple phases \cite{rosati16} in approaching
the ultimate limits in capacity \cite{holevo98, schumacher97}
at the single-photon level.
While in the present work we used long pulses and feedback with a SPD with deadtime of about 35 ns,
the optimized strategies investigated here can be implemented using feed forward, instead of feedback, by
splitting the input state in space and using optical delay lines, as described in Refs. \cite{sych16, becerra11},
allowing for higher repetition rates with much shorter pulses. Together with fast electronic controllers and gated
SPD \cite{bienfang16, zhang15} reaching $>$ GHz speeds, these optimized strategies could allow for high sensitivity
measurements at higher communication rates.
\\
It has been shown \cite{wittmann10} that optimized measurements of two nonorthogonal coherent states based on post-selection can lead to higher secret key rates in QKD than what can be achieved with homodyne measurements and post selection \cite{heid06, sych16}. The optimization methods described here for measurements with definite outcomes may be useful for increasing the rate in QKD protocols with measurements without post-selection \cite{heid06, sych16}. These measurements may lead to higher rates than the heterodyne detection, and could be experimentally investigated with our current setup.
\\
\\
\textbf{Conclusion}
\\Optimized measurements for the discrimination
of multiple nonorthogonal states at the single-photon level can increase the
information transfer in communications and enhance security in
quantum communication. We demonstrate discrimination of multiple
nonorthogonal states below the QNL at the single-photon level under
realistic loss, noise, and system imperfections based on globally optimized
strategies requiring only a few adaptive measurements. These
optimized discrimination measurements can enable low-power
communication with multi-level encoding surpassing the QNL; yield
higher information transfer at low powers; and could have applications
in QKD with multiple coherent states. Our work
makes measurements achieving sensitivities beyond conventional
Gaussian measurements a more realistic alternative to enhance
information transfer beyond what can be achieved with conventional
technologies. We expect that this work will motivate further
theoretical and experimental research in finding optimal
measurements for multiple coherent states and their applications in
quantum information processing and communication.
\\
\\
\textbf{METHODS}\\
\textbf{Preparation of the displacement field.}
The field programmable gate array (FPGA)
provides two digital signals to control the phase of the displacement field $\beta$,
and eight digital signals plus one clock to control its amplitude $|\beta|$, which
corresponds to the optimized displacements $[\beta]$ for a given discrimination strategy.
To control the phase of $\beta$, we use a voltage source together with a 4-to-1 fast multiplexer switch with 210 MHz bandwidth.
The output of the multiplexer switches among the four voltages based on the two digital inputs from the FPGA,
and controls the voltage to PM2 to provide four
possible phase shifts: $\{0,\pi/2,\pi, 3\pi/2\}$ (The phase of the input state is prepared
with PM1 using the same technique.) To control the displacement field amplitude $|\beta|$,
we use the AM and a digital-to-analog converter (DAC) with maximum update rate of 210 MSPS.
The DAC receives eight digital signals from the FPGA and outputs a 1-byte depth analog voltage,
which is sent to the input of the AM. The DAC is updated by a 48 MHz clock provided by the FPGA,
which allows for updating the amplitude of the displacement field as soon as a
new value of $|\beta|$ is available from the FPGA. The total latency
of the circuitry to update the AM input voltage is $<$ 35ns.
%
\\
\\
\noindent
\textbf{ACKNOWLEDGEMENTS}
\\
This work was supported by NSF Grant PHY-1653670 and PHY-1521016.
\\
\\
\\
\noindent
\textbf{REFERENCES}
%

\begin{thebibliography}{10}
\expandafter\ifx\csname url\endcsname\relax
  \def\url#1{\texttt{#1}}\fi
\expandafter\ifx\csname urlprefix\endcsname\relax\def\urlprefix{URL }\fi
\providecommand{\bibinfo}[2]{#2}
\providecommand{\eprint}[2][]{\url{#2}}

\bibitem{helstrom76}
\bibinfo{author}{Helstrom, C.~W.}
\newblock \emph{\bibinfo{title}{Quantum detection and estimation theory,
  Mathematics in Science and Engineering Vol. 123}}
  (\bibinfo{publisher}{Academic Press}, \bibinfo{address}{New York},
  \bibinfo{year}{1976}).

\bibitem{huttner95}
\bibinfo{author}{Huttner, B.}, \bibinfo{author}{Imoto, N.},
  \bibinfo{author}{Gisin, N.} \& \bibinfo{author}{Mor, T.}
\newblock \bibinfo{title}{Quantum cryptography with coherent states}.
\newblock \emph{\bibinfo{journal}{Phys. Rev. A}} \textbf{\bibinfo{volume}{51}},
  \bibinfo{pages}{1863--1869} (\bibinfo{year}{1995}).

\bibitem{bennett92}
\bibinfo{author}{Bennett, C.~H.}
\newblock \bibinfo{title}{Quantum cryptography using any two nonorthogonal
  states}.
\newblock \emph{\bibinfo{journal}{Phys. Rev. Lett.}}
  \textbf{\bibinfo{volume}{68}}, \bibinfo{pages}{3121--3124}
  (\bibinfo{year}{1992}).

\bibitem{grosshans02}
\bibinfo{author}{Grosshans, F.} \& \bibinfo{author}{Grangier, P.}
\newblock \bibinfo{title}{Continuous variable quantum cryptography using
  coherent states}.
\newblock \emph{\bibinfo{journal}{Phys. Rev. Lett.}}
  \textbf{\bibinfo{volume}{88}}, \bibinfo{pages}{057902}
  (\bibinfo{year}{2002}).

\bibitem{gisin02}
\bibinfo{author}{Gisin, N.}, \bibinfo{author}{Ribordy, G.},
  \bibinfo{author}{Tittel, W.} \& \bibinfo{author}{Zbinden, H.}
\newblock \bibinfo{title}{Quantum cryptography}.
\newblock \emph{\bibinfo{journal}{Rev. Mod. Phys.}}
  \textbf{\bibinfo{volume}{74}}, \bibinfo{pages}{145--195}
  (\bibinfo{year}{2002}).

\bibitem{weedbrook12}
\bibinfo{author}{Weedbrook, C.} \emph{et~al.}
\newblock \bibinfo{title}{Gaussian quantum information}.
\newblock \emph{\bibinfo{journal}{Rev. Mod. Phys.}}
  \textbf{\bibinfo{volume}{84}}, \bibinfo{pages}{621--669}
  (\bibinfo{year}{2012}).

\bibitem{sych10}
\bibinfo{author}{Sych, D.} \& \bibinfo{author}{Leuchs, G.}
\newblock \bibinfo{title}{Coherent state quantum key distribution with multi
  letter phase-shift keying}.
\newblock \emph{\bibinfo{journal}{New J. of Phys.}}
  \textbf{\bibinfo{volume}{12}}, \bibinfo{pages}{053019}
  (\bibinfo{year}{2010}).

\bibitem{leverrier09}
\bibinfo{author}{Leverrier, A.} \& \bibinfo{author}{Grangier, P.}
\newblock \bibinfo{title}{Unconditional security proof of long-distance
  continuous-variable quantum key distribution with discrete modulation}.
\newblock \emph{\bibinfo{journal}{Phys. Rev. Lett.}}
  \textbf{\bibinfo{volume}{102}}, \bibinfo{pages}{180504}
  (\bibinfo{year}{2009}).

\bibitem{leverrier11}
\bibinfo{author}{Leverrier, A.} \& \bibinfo{author}{Grangier, P.}
\newblock \bibinfo{title}{Continuous-variable quantum-key-distribution
  protocols with a non-gaussian modulation}.
\newblock \emph{\bibinfo{journal}{Phys. Rev. A}} \textbf{\bibinfo{volume}{83}},
  \bibinfo{pages}{042312} (\bibinfo{year}{2011}).

\bibitem{qi15}
\bibinfo{author}{Qi, B.}, \bibinfo{author}{Lougovski, P.},
  \bibinfo{author}{Pooser, R.}, \bibinfo{author}{Grice, W.} \&
  \bibinfo{author}{Bobrek, M.}
\newblock \bibinfo{title}{Generating the local oscillator ``locally'' in
  continuous-variable quantum key distribution based on coherent detection}.
\newblock \emph{\bibinfo{journal}{Phys. Rev. X}} \textbf{\bibinfo{volume}{5}},
  \bibinfo{pages}{041009} (\bibinfo{year}{2015}).

\bibitem{soh15}
\bibinfo{author}{Soh, D. B.~S.} \emph{et~al.}
\newblock \bibinfo{title}{Self-referenced continuous-variable quantum key
  distribution protocol}.
\newblock \emph{\bibinfo{journal}{Phys. Rev. X}} \textbf{\bibinfo{volume}{5}},
  \bibinfo{pages}{041010} (\bibinfo{year}{2015}).

\bibitem{vanloock06}
\bibinfo{author}{van Loock, P.} \emph{et~al.}
\newblock \bibinfo{title}{Hybrid quantum repeater using bright coherent light}.
\newblock \emph{\bibinfo{journal}{Phys. Rev. Lett.}}
  \textbf{\bibinfo{volume}{96}}, \bibinfo{pages}{240501}
  (\bibinfo{year}{2006}).

\bibitem{vanloock08}
\bibinfo{author}{van Loock, P.}, \bibinfo{author}{L\"utkenhaus, N.},
  \bibinfo{author}{Munro, W.~J.} \& \bibinfo{author}{Nemoto, K.}
\newblock \bibinfo{title}{Quantum repeaters using coherent-state
  communication}.
\newblock \emph{\bibinfo{journal}{Phys. Rev. A}} \textbf{\bibinfo{volume}{78}},
  \bibinfo{pages}{062319} (\bibinfo{year}{2008}).

\bibitem{munro05}
\bibinfo{author}{Munro, W.~J.}, \bibinfo{author}{Nemoto, K.} \&
  \bibinfo{author}{Spiller, T.~P.}
\newblock \bibinfo{title}{Weak nonlinearities: a new route to optical quantum
  computation}.
\newblock \emph{\bibinfo{journal}{New J. of Phys.}}
  \textbf{\bibinfo{volume}{7}}, \bibinfo{pages}{137} (\bibinfo{year}{2005}).

\bibitem{nemoto04}
\bibinfo{author}{Nemoto, K.} \& \bibinfo{author}{Munro, W.~J.}
\newblock \bibinfo{title}{Nearly deterministic linear optical controlled-not
  gate}.
\newblock \emph{\bibinfo{journal}{Phys. Rev. Lett.}}
  \textbf{\bibinfo{volume}{93}}, \bibinfo{pages}{250502}
  (\bibinfo{year}{2004}).

\bibitem{ralph03}
\bibinfo{author}{Ralph, T.~C.}, \bibinfo{author}{Gilchrist, A.},
  \bibinfo{author}{Milburn, G.~J.}, \bibinfo{author}{Munro, W.~J.} \&
  \bibinfo{author}{Glancy, S.}
\newblock \bibinfo{title}{Quantum computation with optical coherent states}.
\newblock \emph{\bibinfo{journal}{Phys. Rev. A}} \textbf{\bibinfo{volume}{68}},
  \bibinfo{pages}{042319} (\bibinfo{year}{2003}).

\bibitem{kilin11}
\bibinfo{author}{Kilin, S.~Y.} \& \bibinfo{author}{Mikhalychev, A.~B.}
\newblock \bibinfo{title}{Optical qudit-type entanglement creation at long
  distances by means of small cross-kerr nonlinearities}.
\newblock \emph{\bibinfo{journal}{Phys. Rev. A}} \textbf{\bibinfo{volume}{83}},
  \bibinfo{pages}{052303} (\bibinfo{year}{2011}).

\bibitem{vanloock11}
\bibinfo{author}{van Loock, P.}
\newblock \bibinfo{title}{Optical hybrid approaches to quantum information}.
\newblock \emph{\bibinfo{journal}{Laser \& Photon. Rev.}}
  \textbf{\bibinfo{volume}{5}}, \bibinfo{pages}{167--200}
  (\bibinfo{year}{2011}).

\bibitem{collins14}
\bibinfo{author}{Collins, R.~J.} \emph{et~al.}
\newblock \bibinfo{title}{Realization of quantum digital signatures without the
  requirement of quantum memory}.
\newblock \emph{\bibinfo{journal}{Phys. Rev. Lett.}}
  \textbf{\bibinfo{volume}{113}}, \bibinfo{pages}{040502}
  (\bibinfo{year}{2014}).

\bibitem{croal16}
\bibinfo{author}{Croal, C.} \emph{et~al.}
\newblock \bibinfo{title}{Free-space quantum signatures using heterodyne
  measurements}.
\newblock \emph{\bibinfo{journal}{Phys. Rev. Lett.}}
  \textbf{\bibinfo{volume}{117}}, \bibinfo{pages}{100503}
  (\bibinfo{year}{2016}).

\bibitem{arrazola14}
\bibinfo{author}{Arrazola, J.~M.} \& \bibinfo{author}{L\"utkenhaus, N.}
\newblock \bibinfo{title}{Quantum communication with coherent states and linear
  optics}.
\newblock \emph{\bibinfo{journal}{Phys. Rev. A}} \textbf{\bibinfo{volume}{90}},
  \bibinfo{pages}{042335} (\bibinfo{year}{2014}).

\bibitem{guan16}
\bibinfo{author}{Guan, J.-Y.} \emph{et~al.}
\newblock \bibinfo{title}{Observation of quantum fingerprinting beating the
  classical limit}.
\newblock \emph{\bibinfo{journal}{Phys. Rev. Lett.}}
  \textbf{\bibinfo{volume}{116}}, \bibinfo{pages}{240502}
  (\bibinfo{year}{2016}).

\bibitem{becerra15}
\bibinfo{author}{Becerra, F.~E.}, \bibinfo{author}{Fan, J.} \&
  \bibinfo{author}{Migdall, A.}
\newblock \bibinfo{title}{Photon number resolution enables quantum receiver for
  realistic coherent optical communications}.
\newblock \emph{\bibinfo{journal}{Nature Photon.}}
  \textbf{\bibinfo{volume}{9}}, \bibinfo{pages}{48–53} (\bibinfo{year}{2015}).

\bibitem{rosati16}
\bibinfo{author}{Rosati, M.}, \bibinfo{author}{Mari, A.} \&
  \bibinfo{author}{Giovannetti, V.}
\newblock \bibinfo{title}{Multiphase hadamard receivers for classical
  communication on lossy bosonic channels}.
\newblock \emph{\bibinfo{journal}{Phys. Rev. A}} \textbf{\bibinfo{volume}{94}},
  \bibinfo{pages}{062325} (\bibinfo{year}{2016}).

\bibitem{lee16}
\bibinfo{author}{Lee, J.}, \bibinfo{author}{Ji, S.-W.}, \bibinfo{author}{Park,
  J.} \& \bibinfo{author}{Nha, H.}
\newblock \bibinfo{title}{Gaussian benchmark for optical communication aiming
  towards ultimate capacity}.
\newblock \emph{\bibinfo{journal}{Phys. Rev. A}} \textbf{\bibinfo{volume}{93}},
  \bibinfo{pages}{050302} (\bibinfo{year}{2016}).

\bibitem{tan15}
\bibinfo{author}{Tan, S.~H.}, \bibinfo{author}{Dutton, Z.},
  \bibinfo{author}{Nair, R.} \& \bibinfo{author}{Guha, S.}
\newblock \bibinfo{title}{Finite codelength analysis of the sequential waveform
  nulling receiver for m-ary psk}.
\newblock In \emph{\bibinfo{booktitle}{2015 IEEE International Symposium on
  Information Theory (ISIT)}}, \bibinfo{pages}{1665--1670}
  (\bibinfo{year}{2015}).

\bibitem{caves94}
\bibinfo{author}{Caves, C.~M.} \& \bibinfo{author}{Drummond, P.~D.}
\newblock \bibinfo{title}{Quantum limits on bosonic communication rates}.
\newblock \emph{\bibinfo{journal}{Rev. Mod. Phys.}}
  \textbf{\bibinfo{volume}{66}}, \bibinfo{pages}{481--537}
  (\bibinfo{year}{1994}).

\bibitem{giovannetti04}
\bibinfo{author}{Giovannetti, V.} \emph{et~al.}
\newblock \bibinfo{title}{Classical capacity of the lossy bosonic channel: The
  exact solution}.
\newblock \emph{\bibinfo{journal}{Phys. Rev. Lett.}}
  \textbf{\bibinfo{volume}{92}}, \bibinfo{pages}{027902}
  (\bibinfo{year}{2004}).

\bibitem{giovannetti14}
\bibinfo{author}{Giovannetti, V.}, \bibinfo{author}{Garcia-Patron, R.},
  \bibinfo{author}{Cerf, N.~J.} \& \bibinfo{author}{Holevo, A.~S.}
\newblock \bibinfo{title}{Ultimate classical communication rates of quantum
  optical channels}.
\newblock \emph{\bibinfo{journal}{Nature Photon.}}
  \textbf{\bibinfo{volume}{8}}, \bibinfo{pages}{796–800}
  (\bibinfo{year}{2014}).

\bibitem{mari14}
\bibinfo{author}{Mari, A.}, \bibinfo{author}{Giovannetti, V.} \&
  \bibinfo{author}{Holevo, A.~S.}
\newblock \bibinfo{title}{Quantum state majorization at the output of bosonic
  gaussian channels}.
\newblock \emph{\bibinfo{journal}{Nature Commun.}}
  \textbf{\bibinfo{volume}{5}}, \bibinfo{pages}{3826} (\bibinfo{year}{2014}).

\bibitem{yuen75}
\bibinfo{author}{Yuen, H.}, \bibinfo{author}{Kennedy, R.} \&
  \bibinfo{author}{Lax, M.}
\newblock \bibinfo{title}{Optimum testing of multiple hypotheses in quantum
  detection theory}.
\newblock \emph{\bibinfo{journal}{IEEE Transactions on Information Theory}}
  \textbf{\bibinfo{volume}{21}}, \bibinfo{pages}{125--134}
  (\bibinfo{year}{1975}).

\bibitem{kennedy72}
\bibinfo{author}{Kennedy, R.~S.}
\newblock \bibinfo{title}{A near-optimum receiver for the binary coherent state
  quantum channel}.
\newblock \bibinfo{note}{Research Laboratory of Electronics, MIT Technical
  Report No. 110 (1972), unpublished}.

\bibitem{taeoka08}
\bibinfo{author}{Takeoka, M.} \& \bibinfo{author}{Sasaki, M.}
\newblock \bibinfo{title}{Discrimination of the binary coherent signal:
  Gaussian-operation limit and simple non-gaussian near-optimal receivers}.
\newblock \emph{\bibinfo{journal}{Phys. Rev. A}} \textbf{\bibinfo{volume}{78}},
  \bibinfo{pages}{022320} (\bibinfo{year}{2008}).

\bibitem{dolinar73}
\bibinfo{author}{Dolinar, S.~J.}
\newblock \bibinfo{title}{An optimum receiver for the binary coherent state
  quantum channel}.
\newblock \bibinfo{note}{Research Laboratory of Electronics, MIT, Quarterly
  Progress Report No. 111 (1973), p. 115.}

\bibitem{cook07}
\bibinfo{author}{Cook, R.~L.}, \bibinfo{author}{Martin, P.~J.} \&
  \bibinfo{author}{Geremia, J.~M.}
\newblock \bibinfo{title}{Optical coherent state discrimination using a
  closed-loop quantum measurement}.
\newblock \emph{\bibinfo{journal}{Nature}} \textbf{\bibinfo{volume}{446}},
  \bibinfo{pages}{774--777} (\bibinfo{year}{2007}).

\bibitem{wittman08}
\bibinfo{author}{Wittmann, C.} \emph{et~al.}
\newblock \bibinfo{title}{Demonstration of near-optimal discrimination of
  optical coherent states}.
\newblock \emph{\bibinfo{journal}{Phys. Rev. Lett.}}
  \textbf{\bibinfo{volume}{101}}, \bibinfo{pages}{210501}
  (\bibinfo{year}{2008}).

\bibitem{tsujino11}
\bibinfo{author}{Tsujino, K.} \emph{et~al.}
\newblock \bibinfo{title}{Quantum receiver beyond the standard quantum limit of
  coherent optical communication}.
\newblock \emph{\bibinfo{journal}{Phys. Rev. Lett.}}
  \textbf{\bibinfo{volume}{106}}, \bibinfo{pages}{250503}
  (\bibinfo{year}{2011}).

\bibitem{bondurant93}
\bibinfo{author}{Bondurant, R.~S.}
\newblock \bibinfo{title}{Near-quantum optimum receivers for the
  phase-quadrature coherent-state channel}.
\newblock \emph{\bibinfo{journal}{Opt. Lett.}} \textbf{\bibinfo{volume}{18}},
  \bibinfo{pages}{1896--1898} (\bibinfo{year}{1993}).

\bibitem{nair12}
\bibinfo{author}{Nair, R.}, \bibinfo{author}{Yen, B.~J.},
  \bibinfo{author}{Guha, S.}, \bibinfo{author}{Shapiro, J.~H.} \&
  \bibinfo{author}{Pirandola, S.}
\newblock \bibinfo{title}{Symmetric $m$-ary phase discrimination using
  quantum-optical probe states}.
\newblock \emph{\bibinfo{journal}{Phys. Rev. A}} \textbf{\bibinfo{volume}{86}},
  \bibinfo{pages}{022306} (\bibinfo{year}{2012}).

\bibitem{izumi12}
\bibinfo{author}{Izumi, S.} \emph{et~al.}
\newblock \bibinfo{title}{Displacement receiver for phase-shift-keyed coherent
  states}.
\newblock \emph{\bibinfo{journal}{Phys. Rev. A}} \textbf{\bibinfo{volume}{86}},
  \bibinfo{pages}{042328} (\bibinfo{year}{2012}).

\bibitem{izumi13}
\bibinfo{author}{Izumi, S.}, \bibinfo{author}{Takeoka, M.},
  \bibinfo{author}{Ema, K.} \& \bibinfo{author}{Sasaki, M.}
\newblock \bibinfo{title}{Quantum receivers with squeezing and
  photon-number-resolving detectors for $m$-ary coherent state discrimination}.
\newblock \emph{\bibinfo{journal}{Phys. Rev. A}} \textbf{\bibinfo{volume}{87}},
  \bibinfo{pages}{042328} (\bibinfo{year}{2013}).

\bibitem{li13}
\bibinfo{author}{Li, K.}, \bibinfo{author}{Zuo, Y.} \& \bibinfo{author}{Zhu,
  B.}
\newblock \bibinfo{title}{Suppressing the errors due to mode mismatch for m
  -ary psk quantum receivers using photon-number-resolving detector}.
\newblock \emph{\bibinfo{journal}{Phot. Tech. Lett., IEEE}}
  \bibinfo{pages}{2182--2184} (\bibinfo{year}{2013}).

\bibitem{muller15}
\bibinfo{author}{M\"{u}ller, C.~R.} \& \bibinfo{author}{Marquardt, C.}
\newblock \bibinfo{title}{A robust quantum receiver for phase shift keyed
  signals}.
\newblock \emph{\bibinfo{journal}{New Journal of Physics}}
  \textbf{\bibinfo{volume}{17}}, \bibinfo{pages}{032003}
  (\bibinfo{year}{2015}).

\bibitem{muller12}
\bibinfo{author}{M\"{u}ller, C.} \emph{et~al.}
\newblock \bibinfo{title}{Quadrature phase shift keying coherent state
  discrimination via a hybrid receiver}.
\newblock \emph{\bibinfo{journal}{New J. of Phys.}}
  \textbf{\bibinfo{volume}{14}}, \bibinfo{pages}{083009}
  (\bibinfo{year}{2012}).

\bibitem{becerra13}
\bibinfo{author}{Becerra, F.~E.} \emph{et~al.}
\newblock \bibinfo{title}{Experimental demonstration of a receiver beating the
  standard quantum limit for multiple nonorthogonal state discrimination}.
\newblock \emph{\bibinfo{journal}{Nature Photon.}}
  \textbf{\bibinfo{volume}{7}}, \bibinfo{pages}{147--152}
  (\bibinfo{year}{2013}).

\bibitem{wittmann10}
\bibinfo{author}{Wittmann, C.}, \bibinfo{author}{Andersen, U.~L.},
  \bibinfo{author}{Takeoka, M.}, \bibinfo{author}{Sych, D.} \&
  \bibinfo{author}{Leuchs, G.}
\newblock \bibinfo{title}{Demonstration of coherent-state discrimination using
  a displacement-controlled photon-number-resolving detector}.
\newblock \emph{\bibinfo{journal}{Phys. Rev. Lett.}}
  \textbf{\bibinfo{volume}{104}}, \bibinfo{pages}{100505}
  (\bibinfo{year}{2010}).

\bibitem{gentile96}
\bibinfo{author}{Gentile, T.~R.}, \bibinfo{author}{Houston, J.~M.} \&
  \bibinfo{author}{Cromer, C.~L.}
\newblock \bibinfo{title}{Realization of a scale of absolute spectral response
  using the national institute of standards and technology high-accuracy
  cryogenic radiometer}.
\newblock \emph{\bibinfo{journal}{Appl. Opt.}} \textbf{\bibinfo{volume}{35}},
  \bibinfo{pages}{4392--4403} (\bibinfo{year}{1996}).

\bibitem{becerra11}
\bibinfo{author}{Becerra, F.~E.} \emph{et~al.}
\newblock \bibinfo{title}{M-ary-state phase-shift-keying discrimination below
  the homodyne limit}.
\newblock \emph{\bibinfo{journal}{Phys. Rev. A}} \textbf{\bibinfo{volume}{84}},
  \bibinfo{pages}{062324} (\bibinfo{year}{2011}).

\bibitem{taeoka14}
\bibinfo{author}{Takeoka, M.} \& \bibinfo{author}{Guha, S.}
\newblock \bibinfo{title}{Capacity of optical communication in loss and noise
  with general quantum gaussian receivers}.
\newblock \emph{\bibinfo{journal}{Phys. Rev. A}} \textbf{\bibinfo{volume}{89}},
  \bibinfo{pages}{042309} (\bibinfo{year}{2014}).

\bibitem{shannon48}
\bibinfo{author}{Shannon, C.~E.}
\newblock \bibinfo{title}{A mathematical theory of communication}.
\newblock \emph{\bibinfo{journal}{The Bell System Technical Journal}}
  \textbf{\bibinfo{volume}{27}}, \bibinfo{pages}{379--423}
  (\bibinfo{year}{1948}).

\bibitem{guha11}
\bibinfo{author}{Guha, S.}
\newblock \bibinfo{title}{Structured optical receivers to attain superadditive
  capacity and the holevo limit}.
\newblock \emph{\bibinfo{journal}{Phys. Rev. Lett.}}
  \textbf{\bibinfo{volume}{106}}, \bibinfo{pages}{240502}
  (\bibinfo{year}{2011}).

\bibitem{jarzyna16}
\bibinfo{author}{Jarzyna, M.}, \bibinfo{author}{Lipi\'{n}ska, V.},
  \bibinfo{author}{Klimek, A.}, \bibinfo{author}{Banaszek, K.} \&
  \bibinfo{author}{Paris, M. G.~A.}
\newblock \bibinfo{title}{Phase noise in collective binary phase shift keying
  with hadamard words}.
\newblock \emph{\bibinfo{journal}{Opt. Express}} \textbf{\bibinfo{volume}{24}},
  \bibinfo{pages}{1693--1698} (\bibinfo{year}{2016}).

\bibitem{holevo98}
\bibinfo{author}{Holevo, A.~S.}
\newblock \bibinfo{title}{The capacity of the quantum channel with general
  signal states}.
\newblock \emph{\bibinfo{journal}{IEEE Trans. on Inf. Theo.}}
  \textbf{\bibinfo{volume}{44}}, \bibinfo{pages}{269--273}
  (\bibinfo{year}{1998}).

\bibitem{schumacher97}
\bibinfo{author}{Schumacher, B.} \& \bibinfo{author}{Westmoreland, M.~D.}
\newblock \bibinfo{title}{Sending classical information via noisy quantum
  channels}.
\newblock \emph{\bibinfo{journal}{Phys. Rev. A}} \textbf{\bibinfo{volume}{56}},
  \bibinfo{pages}{131--138} (\bibinfo{year}{1997}).

\bibitem{sych16}
\bibinfo{author}{Sych, D.} \& \bibinfo{author}{Leuchs, G.}
\newblock \bibinfo{title}{Practical receiver for optimal discrimination of
  binary coherent signals}.
\newblock \emph{\bibinfo{journal}{Phys. Rev. Lett.}}
  \textbf{\bibinfo{volume}{117}}, \bibinfo{pages}{200501}
  (\bibinfo{year}{2016}).

\bibitem{bienfang16}
\bibinfo{author}{Bienfang, J.~C.} \& \bibinfo{author}{Restelli, A.}
\newblock \bibinfo{title}{Characterization of an advanced harmonic subtraction
  single-photon detection system based on an ingaas/inp avalanche diode}.
\newblock \emph{\bibinfo{journal}{Proc. SPIE 9858, Adv. Phot. Count. Techn. X}}
  \textbf{\bibinfo{volume}{98580}}.

\bibitem{zhang15}
\bibinfo{author}{Zhang, J.}, \bibinfo{author}{Itzler, M.~A.},
  \bibinfo{author}{Zbinden, H.} \& \bibinfo{author}{Pan, J.-W.}
\newblock \bibinfo{title}{Advances in ingaas/inp single-photon detector systems
  for quantum communication}.
\newblock \emph{\bibinfo{journal}{Light: Science $\&$ Applications}}
  \textbf{\bibinfo{volume}{4}}, \bibinfo{pages}{e286} (\bibinfo{year}{2015}).

\bibitem{heid06}
\bibinfo{author}{Heid, M.} \& \bibinfo{author}{L\"utkenhaus, N.}
\newblock \bibinfo{title}{Efficiency of coherent-state quantum cryptography in
  the presence of loss: Influence of realistic error correction}.
\newblock \emph{\bibinfo{journal}{Phys. Rev. A}} \textbf{\bibinfo{volume}{73}},
  \bibinfo{pages}{052316} (\bibinfo{year}{2006}).

\end{thebibliography}
%

\end{document}